# Analogue Quantum Gravity in Hyperbolic Metamaterials

Igor I. Smolyaninov [1), Vera N. Smolyaninova [2)

[1) Department of Electrical and Computer Engineering, University of Maryland, College Park, MD 20742, USA;
smoly@umd.edu
[2) Department of Physics, Astronomy, and Geosciences, Towson University, Towson, MD 21252 USA

**Abstract:** It is well known that extraordinary photons in hyperbolic metamaterials may be described as living in an effective Minkowski spacetime, which is defined by the peculiar form of the strongly anisotropic dielectric tensor in these metamaterials. Here we demonstrate that within the scope of this approximation the sound waves in hyperbolic metamaterials look similar to gravitational waves, and therefore the quantized sound waves (phonons) look similar to gravitons. Such an analogue model of quantum gravity looks especially interesting near the phase transitions in hyperbolic metamaterials where it becomes possible to switch quantum gravity effects on and off as a function of metamaterial temperature. We also predict strong enhancement of sonoluminescence in ferrofluid based hyperbolic metamaterials, which looks analogous to particle creation in strong gravitational fields.

## 1. Introduction

Hyperbolic metamaterials are a special class of electromagnetic metamaterials, which exhibit extremely strong anisotropy. These metamaterials exhibit metallic behavior in one direction and dielectric behavior in the orthogonal direction. The original purpose of these metamaterials was to overcome the diffraction limit of optical microscopy [1-2]. However, very soon it was realized that these materials exhibit a very large number of strikingly interesting physical properties, which result from the singular behavior of their photonic density of states [3,4]. In addition to super resolution imaging [1,5,6], it was shown that these materials exhibit enhanced quantum electrodynamic effects [7,8,9] which may be used, for example, in stealth technologies [10]. The transport properties of hyperbolic metamaterials may also be quite unusual, resulting in such effects as thermal hyperconductivity [11], and high Tc superconductivity [12]. It was also pointed out that hyperbolic metamaterials may be used to create very interesting laboratory analogues of gravitational effects [3, 13-16]. While initially it was believed that hyperbolic properties may only be observed in artificial structures, soon it was discovered that many natural materials may also exhibit hyperbolic properties [12,17]. Strikingly enough, even the physical vacuum may potentially exhibit hyperbolic properties [18] when it is subjected to a very strong magnetic field [19].

Essential electromagnetic properties of hyperbolic metamaterials may be understood by considering a non-magnetic uniaxial anisotropic material with dielectric permittivities $\varepsilon_x=\varepsilon_y=\varepsilon_1 >0$ and $\varepsilon_z=\varepsilon_2<0$. Any electromagnetic field propagating in this material may be expressed as a sum of ordinary and extraordinary contributions, each of these being a sum of an arbitrary number of plane waves polarized in the ordinary ($E_z=0$) and extraordinary ($E_z\neq 0$) directions. Let us assume that an extraordinary photon wave function is $\varphi=E_z$ so that the ordinary portion of the electromagnetic field does not contribute to $\varphi$. Maxwell equations in the frequency domain results in the following wave equation for $\varphi_\omega$ if $\varepsilon_1$ and $\varepsilon_2$ are kept constant inside the metamaterial [3]:

$$-\frac{\partial^2 \varphi_\omega}{\varepsilon_1 \partial z^2} + \frac{1}{(-\varepsilon_2)}\left(\frac{\partial^2 \varphi_\omega}{\partial x^2} + \frac{\partial^2 \varphi_\omega}{\partial y^2}\right) = \frac{\omega_0^2}{c^2}\varphi_\omega = \frac{m^{*2} c^2}{\hbar^2}\varphi_\omega \quad , \tag{1}$$

This wave equation coincides with the Klein-Gordon equation for a massive field $\varphi_\omega$ (with an effective mass $m^*$) in a 3D Minkowski spacetime in which one of the spatial coordinates $z = \tau$ behaves as a timelike variable. The metric coefficients $g_{ik}$ of this flat 2+1 dimensional Minkowski spacetime may be defined as [3,13]:

$$g_{00} = -\varepsilon_1 \quad \text{and} \quad g_{11} = g_{22} = -\varepsilon_2 \quad (2)$$

As demonstrated in [13], nonlinear optical Kerr effect may "bend" this 2+1 Minkowski spacetime, resulting in effective gravitational force between the extraordinary photons. It was also predicted that for the effective gravitational constant inside the metamaterial to be positive, negative self-defocusing Kerr medium must be used as a dielectric host of the metamaterial [13].

Artificial hyperbolic metamaterials described by $\varepsilon_x = \varepsilon_y = \varepsilon_1 > 0$ and $\varepsilon_z = \varepsilon_2 < 0$ are typically made out of metal wire array structures, as illustrated in Fig. 1.

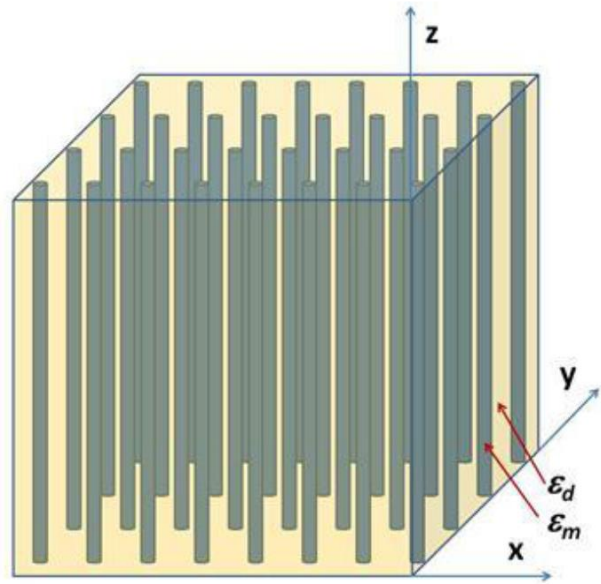

**Figure 1.** Typical geometry of a metal wire array hyperbolic metamaterial.

If the Maxwell-Garnett approximation is applied to such geometry [20], the diagonal components of the permittivity tensor of the metamaterial may be acquired as

$$\varepsilon_1 = \varepsilon_{x,y} = \frac{2f\varepsilon_m\varepsilon_d + (1-f)\varepsilon_d(\varepsilon_d + \varepsilon_m)}{(1-f)(\varepsilon_d + \varepsilon_m) + 2f\varepsilon_d} \approx \frac{1+f}{1-f}\varepsilon_d \;, \qquad \varepsilon_2 = \varepsilon_z = f\varepsilon_m + (1-f)\varepsilon_d \approx f\varepsilon_m \quad (3)$$

where $f$ is the metallic phase volume fraction, and $\varepsilon_m < 0$ and $\varepsilon_d > 0$ are the permittivities of the metal and the dielectric phase, respectively (note that $-\varepsilon_m >> \varepsilon_d$ is typically assumed in the visible and infrared range). On the other hand, as mentioned above, several common natural materials, such as $Al_2O_3$, $ZrSiO_4$, $TiO_2$, *etc.* may also exhibit hyperbolic properties in the long wavelength infrared frequency range [21]. For example, $Al_2O_3$ is naturally hyperbolic in the 19.0-20.4 μm and 23-25 μm frequency bands. Such natural hyperbolic material examples look especially interesting near the phase transitions of the material, since they represent a natural physical situation in which the effective Minkowski spacetime may be "melted" as a function of material temperature $T$ [22]. For example, in the melted state of $Al_2O_3$ the material becomes an isotropic liquid, and the effective Minkowski spacetime experienced by the extraordinary photons becomes an ordinary Euclidean space.

## 2. Methods: Acoustic waves in hyperbolic materials as analogues of gravitational waves

While the relationship between the dielectric permittivity and density of a material may be somewhat complicated, in general the Clausius–Mossotti relation is typically used for this purpose [23]. In the case that the material consists of a mixture of two or more species, the molecular polarizability contribution from each species $\alpha$, indexed by $i$, contributes to the overall dielectric permittivity as follows:

$$\frac{\varepsilon - 1}{\varepsilon + 2} = \sum_i \frac{N_i \alpha_i}{3\varepsilon_0} \quad , \tag{4}$$

where $N_i$ is the molecular concentration of the respective species and $\varepsilon_0$ is the dielectric permittivity of vacuum. As a result, for small deviations of $\varepsilon$ with respect to its average number we may write

$$\Delta\varepsilon = \frac{(\varepsilon+2)^2}{3} \sum_i \frac{\Delta N_i \alpha_i}{3\varepsilon_0} \tag{5}$$

Eqs.(2), (3) and (5) clearly indicate that acoustic waves (the oscillations of molecular concentration $\Delta N$) in a hyperbolic metamaterial act as classical gravitational waves which perturb the effective metric of a flat 2+1 dimensional Minkowski spacetime (see Eq.(2)) experienced by the photons propagating inside the metamaterial. In fact, these "gravitational waves" are known to be quite pronounced in such natural hyperbolic material as sapphire ($Al_2O_3$) due to its very strong piezoelectric behaviour.

Furthermore, the sound waves in hyperbolic metamaterials may be quantized in a straightforward fashion, thus giving rise to the quantum mechanical description of sound waves in terms of phonons. The well-known Hamiltonian for this system is

$$H = \sum \frac{p_i^2}{2m} + \frac{1}{2} m\omega^2 \sum (x_i - x_j)^2 \quad , \tag{6}$$

where $m$ is the mass and $\omega$ is the oscillation frequency of each atom (assuming for simplicity that they are all equal), and $x_i$ and $p_i$ are the position and momentum operators, respectively (the second sum is made over the nearest neighbours). The resulting quantization in momentum space is

$$k_n = \frac{2\pi n}{Na} \quad , \tag{7}$$

where $a$ is the interatomic distance. The harmonic oscillator eigenvalues or energy levels for the mode $\omega_k$ are:

$$E_n = \left(\frac{1}{2} + n\right)\hbar\omega_k \tag{8}$$

While in the $k \to 0$ limit the dispersion relation of phonons is linear, this behaviour changes near $\pi/a$. The picture of "acoustic" and "optical" phonons arises generically if different kinds of atoms are present in the crystalline lattice of the material. For example, if a one-dimensional lattice is made of two types of atoms of mass $m_1$ and $m_2$ connected by a chemical bond which may be characterized by spring constant $K$, two phonon modes result [24]:

$$\omega_\pm = K\left(\frac{1}{m_1} + \frac{1}{m_2}\right) \pm \sqrt{\left(\frac{1}{m_1} + \frac{1}{m_2}\right)^2 - \frac{4\sin^2\frac{ka}{2}}{m_1 m_2}} \quad , \tag{9}$$

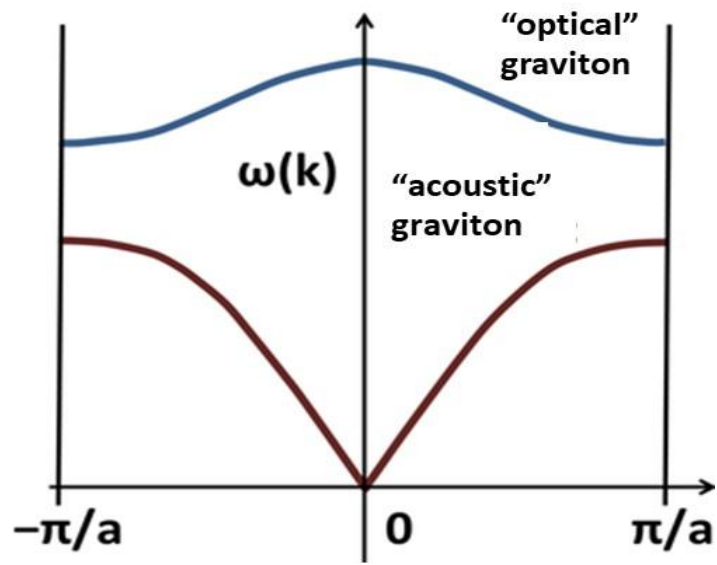

**Figure 2.** Dispersion curves of "acoustic" and "optical" gravitons.

The plus sign corresponds to the "optical mode" in which the two adjacent atoms move against each other, and the minus sign results in the acoustic mode in which they move together. These modes are shown schematically in Fig.2. We should also mention that similar to photons, the dispersion law of phonons in a metamaterial may also be made hyperbolic [25,26], so that both photons and phonons will live in an effective Minkowski spacetime. Quite obviously, the phonons in this picture become the analogues of gravitons. With this conclusion, our analogue model of quantum gravity is basically complete.

**3. Discussion: Analogue quantum gravity effects in hyperbolic metamaterials**

One of the main predictions of various versions of quantum gravity theories is the existence of the minimum length (which typically coincides with the Planck scale) [27]. In the hyperbolic (meta)material version of quantum gravity this minimum length appears naturally due to the finite interatomic distance $a$. However, the minimum length physics actually looks more complicated and quite interesting in such natural ferroelectric hyperbolic materials as $Al_2O_3$ and $BaTiO_3$. Based on Eqs.(1) and (2), the effective length element in a hyperbolic metamaterial equals

$$dl^2 = (-\varepsilon_2)(dx^2 + dy^2) \tag{10}$$

Therefore, the minimum length experienced by the extraordinary photons is

$$l_{min} = \sqrt{-\varepsilon_2}\, a \tag{11}$$

Since $-\varepsilon_2$ is temperature dependent and it may actually diverge near the critical temperature $T_c$ of a ferroelectric phase transition, the analogue quantum gravity effects may become quite pronounced near $T_c$. Indeed, near the transition temperature the dielectric susceptibility, $\chi$, of these materials diverges following the Curie-Weiss law:

$$\chi = (\varepsilon - 1) = \frac{C}{T - T_c} \tag{12}$$

where $C$ is the Curie-Weiss constant of the material.

On the other hand, at the "melting point" of the effective Minkowski spacetime (which was experimentally observed in a ferrofluid-based hyperbolic metamaterial [22] by tuning the $f$ parameter in Eq.(3)) $-\varepsilon_2$ changes sign, and therefore it transitions through the $\varepsilon_2=0$ point as a function of temperature. As a result, the effective "minimum length" inside the metamaterial becomes zero. The possibility of switching quantum gravity effects on and off as a function of metamaterial temperature looks very interesting and attractive. Moreover, since some naturally hyperbolic ferroelectric materials may exhibit quantum criticality [28], truly quantum effects associated with the emergence of the effective Minkowski spacetime at zero temperature may also be studied in the experiment (see also [12]).

Another important prediction of quantum gravity theories are various effects related to particle creation by gravitational fields, such as Hawking radiation, Unruh effect, cosmological particle creation, etc. Let us discuss how these kinds of effects may be experimentally observed and studied within the scope of our model. Based on the analogy between sound and gravitational waves described above, sonoluminescence [29-35] appears to be the most natural choice to search for such effects. Moreover, Eberlein [36,37] already pointed out deep connections between sonoluminescence in liquids and the Unruh effect. According to Eberlein, sonoluminescence occurs when the rapidly moving surface of a microscopic bubble created by ultrasound converts virtual photons into real ones. Like in Unruh effect, the resulting sonoluminescence spectrum appears to be similar to a black-body spectrum. For example, the radiated spectral density within the scope of this model is

$$P(\omega) = 1.16 \frac{(\varepsilon-1)^2}{64\varepsilon} \frac{\hbar}{c^4 \gamma} \left(R_0^2 - R_{min}^2\right)^2 \omega^3 e^{-2\gamma\omega} \quad , \tag{13}$$

where $R_0$ and $R_{min}$ correspond to changes in bubble radius and $\gamma$ describes the timescale of the bubble collapse (see Eq.(4.12) from [36]). This expression is indeed proportional to the energy density of thermal radiation given by the usual Planck expression

$$\frac{dU}{d\omega} = \frac{1}{\pi^2 c^3} \frac{\hbar \omega^3}{e^{\hbar\omega/kT}-1} \qquad (14)$$

if we assume that $T \sim 1/\gamma$.

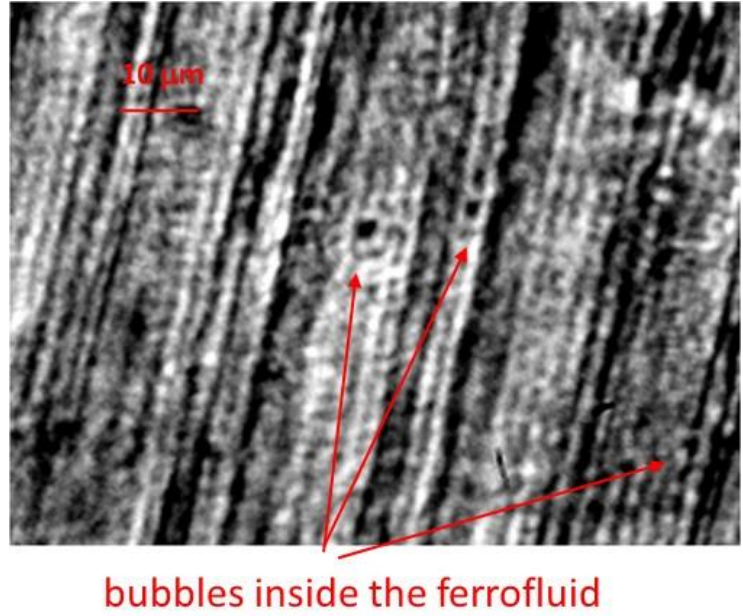

**Figure 3.** Photo of a ferrofluid-based self-assembled hyperbolic metamaterial [22]. The cobalt nanoparticle chains are formed inside the ferrofluid after application of external magnetic field. Several micron-scale bubbles inside the ferrofluid are indicated by arrows.

As was recently predicted in [38], the Unruh effect is supposed to be strongly enhanced inside hyperbolic metamaterials, so that many orders of magnitude smaller accelerations may be used to observe the Unruh radiation. Following this prediction, we may demonstrate that sonoluminescence in ferrofluid-based hyperbolic metamaterials [14,22] will be strongly enhanced too. Let us consider a microscale sonoluminescent bubble inside the ferrofluid, as illustrated in Fig.3. The enhancement of Unruh effect in hyperbolic metamaterials originates from the modification of the conventional Planck expression for the energy density of thermal radiation (Eq.(14)) due to the huge enhancement of the photonic density of states $\rho(\omega)$ inside the metamaterial [3,7]. In particular, for the nanowire array metamaterial design shown in Fig.1 this enhancement factor equals

$$\frac{S_T}{S_T^{(0)}} \approx \frac{5}{16\pi^2} \left( \frac{k_{max}^2}{k_T k_p} \right)^2 \qquad , \qquad (15)$$

where $S_T$ is the energy flux along the symmetry axis of the metamaterial and $S^{(0)}_T$ is the usual Planck value for the energy flux (see Eq.(9) from [38]). The characteristic $k$-vectors in Eq.(15) are $k_{max} \sim 1/a$ (defined by the structural parameter $a$ of the metamaterial), $k_T = k_B T/\hbar c$ is the typical thermal momentum, and $k_p$ is the typical "plasma momentum" [39] of the metamaterial, which is defined as

$$k_p = \sqrt{\frac{4\pi N}{m^*}} \frac{e}{c} \qquad , \qquad (16)$$

where $N$ and $m^*$ are the free charge carrier density in the metamaterial and their effective mass, respectively. Since in a typical metamaterial $k_{max}$ is several orders of magnitude larger than $k_T$ and $k_p$, [40] the enhancement factor defined by Eq.(15) may reach up to ten orders of magnitude (recall that $a \sim 1$nm in natural hyperbolic materials). As a result, a similar strong enhancement may be expected for the radiated spectral density of sonoluminescence defined by Eq.(13).

To summarize, it appears that several quantum gravity effects find interesting analogues in artificial and natural hyperbolic metamaterials. While the described analogies are obviously not perfect, the fact that these effects may be studied in the lab [14,22] make them a very useful tool to develop more intuition on the actual inner working of quantum gravity.

## 4. Conclusions

In conclusion, based on the fact that extraordinary photons in hyperbolic metamaterials [41-43] may be described as living in an effective Minkowski spacetime (which is defined by the peculiar form of the strongly anisotropic dielectric tensor in these metamaterials) we have demonstrated that sound waves in hyperbolic metamaterials look similar to gravitational waves [44-47]. As a result, within the scope of this model the quantized sound waves (phonons) look similar to gravitons [48-50]. Such an analogue model of quantum gravity looks especially interesting near the phase transitions in hyperbolic metamaterials where it becomes possible to switch quantum gravity effects on and off at will as a function of metamaterial temperature. We also predicted strong enhancement of sonoluminescence in ferrofluid based hyperbolic metamaterials, which looks analogous to such important quantum gravity effects as particle creation in gravitational fields [51-61].